\documentclass[pra, a4paper, twocolumn, showpacs,superscriptaddress]{revtex4}
\usepackage{amsmath}
\usepackage{amsfonts}
\usepackage{graphicx}
\usepackage{bbm}
\usepackage{longtable}
\begin{document}

\title{Joint measurability, steering and entropic uncertainty} 
\author{H. S. Karthik}\email{karthik@rri.res.in}\affiliation{Raman Research Institute, Bangalore 560 080, India} 
\author{A. R. Usha Devi}\email{arutth@rediffmail.com} 
\affiliation{Department of Physics, Bangalore University, 
Bangalore-560 056, India}
\affiliation{Inspire Institute Inc., Alexandria, Virginia, 22303, USA.}
\author{A. K. Rajagopal} \email{attipat.rajagopal@gmail.com}
\affiliation{Inspire Institute Inc., Alexandria, Virginia, 22303, USA.}
\affiliation{Institute of Mathematical Sciences, C.I.T. Campus, Taramani, Chennai, 600113, India}
\affiliation{Harish-Chandra Research Institute, Chhatnag Road, Jhunsi, Allahabad 211 019, India.}
\date{\today}
\begin{abstract} 
 There has been a surge of research activity recently on the role of joint measurability of unsharp observables on non-local features viz., violation of Bell inequality and EPR steering. Here, we investigate the entropic uncertainty relation for a pair of non-commuting observables (of Alice's system), when an entangled quantum memory of Bob is restricted to record outcomes of jointly measurable POVMs. We show that with this imposed constraint of  joint measurability at Bob's end, the  entropic uncertainties associated with Alice's measurement outcomes -- conditioned by the results registered at Bob's end --  obey an entropic steering inequality. Thus, Bob's non-steerability is intrinsically linked with his inability in predicting the outcomes of
Alice's pair of non-commuting observables with better precision, even when they share an entangled state. As a further consequence, we prove that in the joint measurability regime,  the quantum advantage envisaged for the construction of security proofs in quantum key distribution is lost. 
    
\end{abstract}
\maketitle
\section{Introduction}
In the classical domain physical observables commute with each other and  they can all be jointly measured. In contrast, measurements of observables, which do not commute are usually declared to be {\em incompatible} in the quantum scenario. However, the notion of {\em compatibility} of measurements is captured entirely by {\em commutativity} of the observables if one restricts only to {\em sharp} projective valued (PV) measurements. In an extended framework,  which include measurements of {\em unsharp} generalized observables, comprised of  positive operator valued measures (POVM),   the concept of {\em joint measurability} gets delinked from that of commutativity~\cite{Busch, Lahti1, Barnett, Son, Stano,  BuschLahti, Wolf, Yu,  Wolf2, LSW}. Though non-commuting observables do not admit simultaneous {\em sharp} values through their corresponding PV measurements, it is possible to assign {\em unsharp} values  jointly to compatible positive operator valued (POV) observables. Active research efforts are dedicated~\cite{Busch, Barnett, Son, Stano, Wolf, Kar, Kunjwal, Brunner, Guhne} to explore clear, operationally significant criteria  of {\em joint measurability} for  two or more POVMs  and also to identify that incompatible measurements, which cannot be implemented jointly, are necessary to bring out {\em non-classical} features.  In this context, it has already been recognized~\cite{Busch, Barnett, Son, Stano, Wolf, Kar, Brunner, Guhne} that if one merely confines to local compatible  POVMs  on  parts of an entangled quantum system, it is not possible to witness non-local quantum features like steering (the ability to non-locally alter the states of one part of a composite system by performing measurements on another spatially separated part~\cite{Sch}) and violation of Bell inequality. More specifically,  incompatible measurements are instrumental in bringing to surface the violations of various no-go theorems in the quantum world. 

In this work, we investigate the entropic uncertainty relation associated with Alice's measurements of a  pair of non-commuting discrete  observables with $d$ outcomes, in the presence of Bob's quantum memory~\cite{Berta} -- by restricting to compatible (jointly measurable) POVMs at Bob's end. We first establish  that the sum of entropies of Alice's measurement results, when conditioned by the outcomes of compatible {\em unsharp} POVMs recorded in  Bob's quantum memory, is constrained to {\em obey} an entropic steering inequality derived in Ref.~\cite{Howell, Walborn}. This essentially brings out the intrinsic equivalence between  the violation of an entropic steering inequality and the possiblity of reducing the entropic uncertainty bound of a pair of non-commuting observables with the help of an entangled quantum memory. And as violation of a steering inequality  requires~\cite{Brunner, Guhne} that (i) the parties share a steerable entangled state and  also  that (ii) the measurements by one of the  parties (Bob)~\cite{note3} is incompatible, it becomes evident that information stored in Bob's entangled quantum memory is of no use in reducing the uncertainty of Alice's pair of non-commuting observables, when Bob can  measure only compatible POVMs. Consequently, we prove that the quantum advantage  for the construction of security proofs in quantum key distribution (QKD)~\cite{Berta} is lost in the joint measurability regime.

The structure of the paper is organized as follows. In Sec.~II we  give an overview of generalized POV observables and  their joint measurability. Entropic uncertainty relation for Alice's  pair of discrete observables in the presence of Bob's quantum memory is discussed in Sec.~III. It is  shown that when Bob is  restricted to employ  only jointly measurable POVMs, it is not possible to achieve enhanced precision for predicting  Alice's measurement outcomes, even if  entangled state is shared between them. Implications of this identification on security proofs in QKD is also outlined. Section IV contains concluding remarks.

\section{Joint measurability} 

We begin by giving an outline of generalized measurement of observables in terms of POVMs. A POVM is a set $\mathbbm{E}=\{E(x)\}$ of positive self-adjoint operators $0\leq E(x)\leq 1$, called {\em effects}, satisfying  $\sum_x E(x)=\mathbbm{1}$; $\mathbbm{1}$ denotes the identity operator. When a quantum system is prepared in the state $\rho$, measurement of  $\mathbbm{E}$ gives an outcome  $x$  with probability $p(x)={\rm Tr}[\rho\, E(x)]$. If $\{E(x)\}$ is a set of complete, orthogonal projectors, then the measurement reduces to the special case of PV measurement.  
 
Let us consider a collection of POV observables  $\mathbbm{E}_i=\{ E_i(x_i)\}$. They are  jointly measurable if there exists a {\em grand} POVM $\mathbbm{G}=\{G(\lambda);\ 0\leq G(\lambda)\leq \mathbbm{1},\, \sum_\lambda\, G(\lambda)=\mathbbm{1}\}$ from which the observables $\mathbbm{E}_i$ can be constructed as follows. Suppose a measurement of the generalized observable $\mathbbm{G}$ is carried out in a state $\rho$ and the probability of obtaining the outcome $\lambda$ is denoted by $p(\lambda)={\rm Tr} [\rho\, G(\lambda)]$. If the elements of the  POVMs  $\mathbbm{E}_i=\{E_i(x_i)\}$ can be constructed as {\em marginals} of the {\em grand} POVM $\mathbbm{G}=\{\,G(\lambda),\ \ \lambda=\{x_1,x_2,\ldots\}\, \}$ i.e.,  
$E_1(x_1)=\sum_{x_2,x_3,\ldots}\, G(x_1,x_2,\ldots)$, $E_2(x_2)=\sum_{x_1,x_3,\ldots}\, G(x_1,x_2,\ldots)$ and so on, the set $\{\mathbbm{E}_i\}$ of POVMs is jointly measurable~\cite{Busch}. 

In general, if the effects $E_{i}(x_i)$ can be constructed in terms of $G(\lambda)$ as~\cite{Guhne, Foun},       
\begin{equation}
\label{masterG}
E_i(x_i)= \sum_{\lambda} p(x_i\vert i, \lambda)\, G(\lambda)\ \ \  \forall \ \  i
\end{equation}
where $0\leq p(x_i\vert i, \lambda)\leq 1$ are positive numbers satisfying $\sum_{x_i}\, p(x_i\vert i, \lambda)=1$, then the POVMs $\mathbbm{E}_i$ are jointly measurable~\cite{note}.  
For all jointly measurable  POVMs, the probability $p(x_i\vert i)$ of the outcome $x_i$ in the measurement of $\mathbbm{E}_i$ can be post processed based on the results of measurement of  the grand POV observale $\mathbbm{G}$:  
\begin{equation}
p(x_i\vert i)={\rm Tr}[\rho\, E_i(x_i)]=\sum_\lambda\, p(\lambda)\, p(x_i\vert i,\lambda). 
\end{equation}
More specifically, measurements of compatible POVMs $\mathbbm{E}_i$ can be interpreted in terms of a single {\em grand} POVM $\mathbbm{G}$  (i.e., given the positive numbers $p(x_i\vert i,\lambda)$, one can construct the probabilities of measuring compatible POVMs $\mathbbm{E}_i$ solely based on the results of measurement of $\mathbbm{G}$).   
An important feature that gets highlighted here is that the generalized POV observables are jointly  measurable even if they do not commute with each other. 

Reconciling to joint measurability within quantum theory results in subsequent manifestation of classical features~\cite{Guhne}. In particular, as  measurement of a single {\em grand} POVM can be used to construct results of measurements of all compatible POVMs, joint measurability entails a  joint probability distribution for all compatible observables (though for {\em unsharp} values of the observables) in {\em every} quantum state. Existence of joint probabilities in turn implies that  the set of all  Bell inequalities are satisfied~\cite{Fine}, when only compatible measurements are employed.  Wolf et. al.~\cite{Wolf} have shown that incompatible measurements of a pair of POVMs with dichotomic outcomes are necessary and sufficient for the violation of Clauser-Horne-Shimony-Holt (CHSH) Bell inequality. Further, Quintino et. al.~\cite{Brunner} and Uola et. al.~\cite{Guhne} have established a more general result that a set of  POVMs (with arbitrarily many outcomes) are not jointly measurable if and only if they are useful for  non-local quantum steering. It is of interest to explore  the limitations imposed  by joint measurability on quantum information tasks. 
In the following, we study the implications of joint measurability on entropic uncertainty relation in the presence of quantum memory.

\section{ Entropic uncertainty relation in the presence of quantum memory} 

The Shannon entropies $H(\mathbbm{X})=-\sum_x \, p(x) \log_2 p(x)$, 
$H(\mathbbm{Z})=-\sum_z \, p(z) \log_2 p(z)$,  associated with the probabilities $p(x)={\rm Tr}\,[\rho\, E_{\mathbbm{X}}(x)]$, $p(z)={\rm Tr}\,[\rho\, E_{\mathbbm{Z}}(z)]$ of  measurement outcomes $x,\, z$ of a pair of POV observables $\mathbbm{X}\equiv\{E_{\mathbbm{X}}(x)\vert\, 0\leq E_{\mathbbm{X}}\leq \mathbbm{1};\ \sum_x\, E_{\mathbbm{X}}=\mathbbm{1}\} ,\ \mathbbm{Z}\equiv\{E_{\mathbbm{Z}}(z)\vert\, 0\leq E_{\mathbbm{Z}}\leq \mathbbm{1};\ \sum_z\, E_{\mathbbm{Z}}=\mathbbm{1}\}$, quantify the uncertainties of predicting the measurement outcomes in a quantum state $\rho$.  Trade-off between the entropies of  observables $\mathbbm{X}$ and 
$\mathbbm{Z}$ in a finite level quantum system is quantified by the entropic uncertainty relation~\cite{MU, KP}: 
\begin{equation} 
\label{kpbound}
H(\mathbbm{X})+H(\mathbbm{Z})\geq - 2\log_2\, {\cal C}(\mathbbm{X},\mathbbm{Z}),
\end{equation}
where ${\cal C}(\mathbbm{X},\mathbbm{Z})={\rm max}_{x,z}\  \vert\vert \sqrt{E_{\mathbbm{X}}(x)}\, \sqrt{E_{\mathbbm{Z}}(z)}\vert\vert$. (Here, $\vert\vert A\vert\vert={\rm Tr}[\sqrt{A^\dag\, A}]$).

\begin{figure}
\includegraphics*[width=3.4in,keepaspectratio]{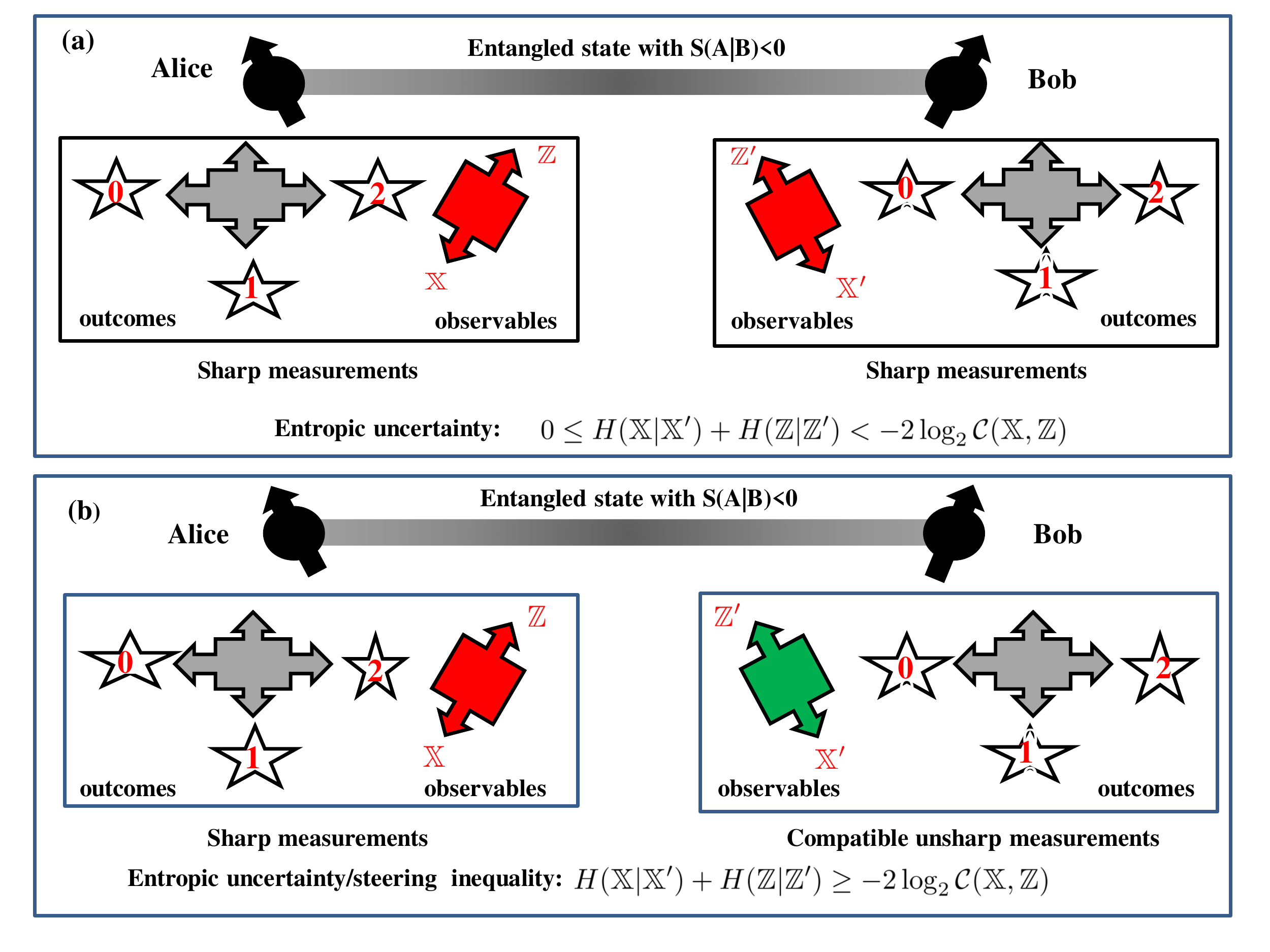}
\caption{(Color online) Alice and Bob decide on a pair of non-commuting observables $\mathbbm{X},\, \mathbbm{Z}$. Bob prepares an entangled state $\rho_{AB}$ and sends the subsystem $A$ to Alice. Alice measures $\mathbbm{X}$ or  $\mathbbm{Z}$ randomly and conveys her choice to Bob. At his end, Bob measures $\mathbbm{X}'$ or  $\mathbbm{Z}'$ and predicts Alice's outcomes. (a) Alice and Bob both perform sharp measurements. In this case, Bob can predict Alice's outcomes with an enhanced precision, as the entropic uncertainty bound (see (\ref{BobM})) can be smaller than $- 2\log_2 C(\mathbbm{X},\mathbbm{Z})$, when the conditional von Neumann entropy  $S(A\vert B)$ of the entangled state $\rho_{AB}$ is negative. (b) Alice performs sharp measurements of the chosen observables $\mathbbm{X}$ or  $\mathbbm{Z}$, while Bob correspondingly records outcomes of compatible unsharp measurements of $\mathbbm{X}'$ or $\mathbbm{Z}'$. In the joint measurability range of $\mathbbm{X}',\, \mathbbm{Z}'$, Bob's quantum memory fails to predict Alices outcomes more precisely because the sum of entropies $H(\mathbbm{X}\vert \mathbbm{X}' )$, $H(\mathbbm{Z}\vert \mathbbm{Z}' )$ is constrained to obey an entropic steering inequality: $H(\mathbbm{X}\vert \mathbbm{X}' )+H(\mathbbm{Z}\geq - 2\log_2 C(\mathbbm{X},\mathbbm{Z})$.}
\end{figure}

Consider the following uncertainty game~\cite{Berta}: two players Alice and Bob agree to measure a pair of observables $\mathbbm{X}$ and $\mathbbm{Z}$. Bob prepares a quantum state of his choice and sends it to Alice. Alice  measures $\mathbbm{X}$ or $\mathbbm{Z}$ randomly and communicates her choice of measurements to Bob. To win the game, Bob's initial preparation of the quantum state should be such that he is able to predict Alice's measurement outcomes of the chosen pair of observables $\mathbbm{X}$ or $\mathbbm{Z}$ with as much precision as possible, when Alice announces which of the pair of observables is measured. In other words,  Bob's task is to minimize the uncertainties in the measurements of a pair of observables $\mathbbm{X}, \ \mathbbm{Z}$ that were agreed upon initially, with the help of an optimal quantum state. The uncertainties of $\mathbbm{X}, \ \mathbbm{Z}$ are bounded as in (\ref{kpbound}), when Bob has only classical information about the state. On the other hand, with the help of a quantum memory (where Bob prepares an entangled state and sends one part of the state to Alice) Bob can beat the uncertainty bound of (\ref{kpbound}).

The entropic uncertainty relation, when Bob possesses a quantum memory, was put forth by Berta et. al.,~\cite{Berta}:  
\begin{equation} 
\label{bertabound}
S(\mathbbm{X}\vert B)+ S(\mathbbm{Z}\vert B)\geq -2\log_2 {\cal C}(\mathbbm{X},\mathbbm{Z}) +S(A\vert B),
\end{equation}
where $S(\mathbbm{X}\vert B)=S\left(\rho^{(\mathbbm{X})}_{AB}\right)-S(\rho_B),$  $S(\mathbbm{Z}\vert B)=
S\left(\rho^{(\mathbbm{Z})}_{AB}\right)-S(\rho_B)$ are the conditional von Neumann entropies of the post measured states 
$\rho^{(\mathbbm{X})}_{AB}=\sum_x\, \vert x\rangle\langle x\vert \otimes \rho^{(x)}_B$ with $\rho^{(x)}_B={\rm Tr}_A[\rho_{AB} (E_{\mathbbm{X}}(x) \otimes \mathbbm{1}_B)]$ and  $\rho^{(\mathbbm{Z})}_{AB}=\sum_z\, \vert z\rangle\langle z\vert \otimes \, \rho^{(z)}_B$ with  $\rho^{(z)}_B={\rm Tr}_A[\rho_{AB} (E_{\mathbbm{Z}}(z) \otimes \mathbbm{1}_B)]$,  obtained after Alice's measurements of $\mathbbm{X},\, \mathbbm{Z}$  on her system. (Here, the measurement outcomes of the effects $E_{\mathbbm{X}}(x)$ ($E_{\mathbbm{Z}}(z)$) are encoded in an orthonormal basis $\{\vert x\rangle\}$ ($\{\vert z\rangle\}$) and the probability of measurement outcome $x$ ($z$) is given by $p(x)={\rm Tr}[\rho^{(x)}_B]$  ($p(z)={\rm Tr}[\rho^{(z)}_B]$);  $S(A\vert B)=S(\rho_{AB})-S(\rho_B)$ is the conditional  von Neumann entropy of the state $\rho_{AB}$). 

When Alice's system is in a maximally entangled state with Bob's quantum memory, the second term on the right hand side of (\ref{bertabound}) takes the value $S(A\vert B)=-\log_2 d$ and as $- 2\log_2 C(\mathbbm{X},\mathbbm{Z})\leq \log_2 d$ (which can be realized when Alice employs pairs of unbiased projective measurements~\cite{Wehner}),  a trivial lower bound of zero is obtained in the entropic uncertainty relation.  In other words, by sharing  an entangled state with Alice, Bob can beat the uncertainty bound given by (\ref{kpbound}) and can predict the outcomes of a pair of observables $\mathbbm{X}$, $\mathbbm{Z}$ with improved precision by performing suitable measurements on his part of the state. 

Let us denote $\mathbbm{X}'$ or $\mathbbm{Z}'$ as the POVMs which Bob choses to measure, when Alice announces her choice of measurements of the observables $\mathbbm{X}$ or $\mathbbm{Z}$. The uncertainty relation (\ref{bertabound}) can be recast in terms of the conditional entropies~\cite{note2} $H(\mathbbm{X}\vert \mathbbm{X}' ),\ H(\mathbbm{Z}\vert \mathbbm{Z}' )$ of Alice's measurement outcomes of the observables $\mathbbm{X}$, $\mathbbm{Z}$,  conditioned  by Bob's measurements of $\mathbbm{X}'$, $\mathbbm{Z}'$. As measurements always increase entropy i.e.,  $H(\mathbbm{X}\vert \mathbbm{X}')\geq S(\mathbbm{X}\vert B)$, $H(\mathbbm{Z}\vert \mathbbm{Z}')\geq S(\mathbbm{Z}\vert B)$,  the entropic uncertainty relation in the presence of quantum memory can be expressed in the form~\cite{Berta}
\begin{equation} 
\label{BobM}
H(\mathbbm{X}\vert \mathbbm{X}' )+ H(\mathbbm{Z}\vert \mathbbm{Z}')\geq -2\log_2 {\cal C}(\mathbbm{X},\mathbbm{Z}) + S(A\vert B). 
\end{equation}

On the other hand,  the conditional entropies $H(\mathbbm{X}\vert \mathbbm{X}' )$, $H(\mathbbm{Z}\vert \mathbbm{Z}')$ are constrained to obey the {\em entropic steering inequality}~\cite{Howell, Walborn},
\begin{equation}
\label{esr} 
H(\mathbbm{X}\vert \mathbbm{X}' )+ H(\mathbbm{Z}\vert \mathbbm{Z}') \geq -2\log_2 {\cal C}(\mathbbm{X},\mathbbm{Z})
\end{equation}
if Bob is unable to remotely steer Alice's state by his local measurements. And, as has been proved recently~\cite{Brunner, Guhne}, measurements at Bob's end can result in the violation of any steering inequality if and only if they are incompatible (in addition that the state shared between Alice and Bob is entangled so as to be steerable). In other words,  the entropic inequality (\ref{esr}) can never be violated if Bob's measurements $\mathbbm{X}'$, $\mathbbm{Z}'$ are compatible. Violation of the steering inequality (\ref{esr}) would in turn correspond to a reduced bound in the entropic uncertainty relation (\ref{BobM}) in the presence of quantum memory (reduction in the bound is realized 
 when Alice and Bob share an entangled state with $S(A\vert B)<0$).  If Bob is constrained to  perform compatible measurements on his system,  he cannot beat the uncertainty bound of (\ref{kpbound}) and win the {\em uncertainty game} by predicting the outcomes as precisely as possible, even when he shares a maximally entangled state with Alice (See Fig.~1).

\subsection{An example}

We illustrate the entropic uncertainty relation (\ref{bertabound}) for  a pair of  qubit observables $\mathbbm{X}=\vert 0\rangle\langle 1\vert+\vert 1\rangle\langle 0\vert$ and $\mathbbm{Z}=\vert 0\rangle\langle 0\vert-\vert 1\rangle\langle 1\vert$, when Alice and Bob share a maximally entangled two-qubit state 
$\vert \psi\rangle_{AB}=\frac{1}{\sqrt{2}}\, \left(\vert 0_A,\ 1_B\rangle - \vert 1_A,\ 0_B\rangle\right)$. Alice performs one of the {\em sharp} PV measurements 
\begin{eqnarray}
\label{sharpxz}
\Pi_{\mathbbm{X}}(x)&=&\frac{1}{2}\left(\mathbbm{1}+x\, \mathbbm{X}\right),\ x=\pm 1,\nonumber \\
\Pi_{\mathbbm{Z}}(z)&=&\frac{1}{2}\left(\mathbbm{1}+z\, \mathbbm{Z} \right),\ z=\pm 1, 
\end{eqnarray}
of  the observables $\mathbbm{X}$ or $\mathbbm{Z}$ randomly on her qubit and announces her choice of measurement, while Bob tries to predict Alice's outcomes by  performing {\em unsharp} compatible measurements of 
$\mathbbm{X}'=\{ E_{\mathbbm{X}'}(x'),\ x'=\pm 1\}$ or $\mathbbm{Z}'=\{ E_{\mathbbm{Z}'}(z'),\ z'=\pm 1\}$ on his qubit. The effects $E_{\mathbbm{X}'}(x'), E_{\mathbbm{Z}'}(z')$ constituting the binary {\em unsharp} qubit observables  $\mathbbm{X}',\ \mathbbm{Z}'$ are given by,  
\begin{eqnarray}
\label{unsharpxz}
 E_{\mathbbm{X}'}(x')&=&\frac{1}{2}\left(\mathbbm{1}+ \eta\, x'\,\mathbbm{X}' \right),\nonumber \\
E_{\mathbbm{Z}'}(z')&=&\frac{1}{2}\left(\mathbbm{1}+ \eta\, z'\,\mathbbm{Z}' \right),
\end{eqnarray} 
where $x'$, $z'$ are the measurement outcomes  and  $0\leq \eta\leq 1$ denotes the {\em unsharpness}  of  the fuzzy measurements. Clearly, when $\eta=1$, the POVM elements $E_{\mathbbm{X}'}(x')$, $E_{\mathbbm{Z}'}(z')$ reduce to their corresponding {\em sharp} PV versions (see  (\ref{sharpxz})) $\Pi_{\mathbbm{X}'}(x'), \Pi_{\mathbbm{Z}'}(z')$.

The joint probabilities $p(x,x')$ (or $p(z,z')$ of Alice's {\em sharp} outcome $x$ (or $z$) and Bob's {\em unsharp} outcome $x'$ (or $z'$), when they both choose to measure the same observable $\mathbbm{X}$ (or $\mathbbm{Z}$) at their ends,  is obtained to be, 
\begin{eqnarray}
p(x,x')&=&\langle \psi_{AB}\vert \Pi_{\mathbbm{X}}(x) \otimes E_{\mathbbm{X}}(x')\vert \psi_{AB}\rangle \nonumber \\  
&=& \frac{1}{4}\left(1 - \eta\, x\, x')\right)  \nonumber \\
p(z,z')&=&\langle \psi_{AB}\vert \Pi_{\mathbbm{Z}}(z) \otimes E_{\mathbbm{Z}}(z')\vert \psi_{AB}\rangle \nonumber \\
&=& \frac{1}{4}\left(1 - \eta\, z\, z')\right)  
\end{eqnarray}
While the right-hand side of  the entropic uncertainty relation (\ref{BobM}) reduces to zero in this case, the left-hand side can be simplified~ (see \cite{note2}) to obtain,   
\begin{widetext}
\begin{eqnarray}
\label{lhs1}
H(\mathbbm{X}\vert\mathbbm{X}')+ H(\mathbbm{Z}\vert\mathbbm{Z}')&=& -\sum_{x,x'=\pm 1}\, p(x,x') \log_2 p(x\vert x') %\nonumber \\ & & \ 
-\sum_{z,z'=\pm 1}\, p(z,z') \log_2 p(z\vert z')\nonumber \\
&=& 2\, H[(1+\eta)/2]
\end{eqnarray}  
\end{widetext}
where $H(p)=-p\, \log_2 p-(1-p)\, \log_2 (1-p)$ is the binary entropy. As the binary entropy function $H[(1+\eta)/2]=0$ only when $\eta=1$, the trivial lower bound  of the uncertainty relation (\ref{BobM})  can be reached  if Bob too performs {\em sharp} PV measurements of the observables $\mathbbm{X}$ and $\mathbbm{Z}$ at his end. In other words, Bob can predict the outcomes of Alice's measurements of $\mathbbm{X}$ and  $\mathbbm{Z}$ precisely when he employs {\em sharp} PV measurements of the same observables. But {\em sharp}   measurements of $\mathbbm{X}$ and  $\mathbbm{Z}$ are not compatible. The joint measurability of the unsharp POVMs   $\mathbbm{X}=\{E_{\mathbbm{X}}(x')\}$ and $\mathbbm{Z}=\{E_{\mathbbm{Z}}(z')\}$ sets the limitation~\cite{Busch, Stano}\ \  $\eta\leq 1/\sqrt{2}$ on the {\em unsharpness} parameter. 

If Bob confines only to  the joint measurability range $0\leq \eta\leq 1/\sqrt{2}$ in the measurement of   $\mathbbm{X}=\{E_{\mathbbm{X}}(x')\}$ and $\mathbbm{Z}=\{E_{\mathbbm{Z}}(z')\}$, the entropic steering inequality (\ref{esr})
\begin{equation}
\label{esr2}
H(\mathbbm{X}\vert\mathbbm{X}')+ H(\mathbbm{Z}\vert\mathbbm{Z}')\geq 1     
\end{equation} 
is always satisfied~\cite{note4}.  In turn, it implies that  Bob cannot beat the entropic uncertainty bound of (\ref{kpbound}) --  even with the help of an entangled state he shares with Alice --   if he is constrained to employ jointly measurable POVMs. 

The result demonstrated here in the specific example of  $d=2$ (qubits) holds in principle for any $d$ dimensional POVMs. An illustration in the $d$ dimensional example, however, requires that the compatibility/incompatibility range of the unsharpness parameter $\eta$ is known. However, optimal values of the unsharpness parameter ($\eta$) of  a set of  POVMs is known only for qubits.

\subsection{Joint measurability and QKD}

The entropic uncertainty relation in the presence of quantum memory (\ref{bertabound}) provides a quantification for the connection between entanglement and uncertainty. Moreover, it has been shown~\cite{Berta} to be useful to derive a lower bound on the secret key rate that can be generated by Alice and Bob in QKD  against collective attacks by an adversary Eve. Subsequently more tighter finite-key bound on  discrete variable QKD has been derived based on generalized uncertainty relations  for smooth min- max- entropies~\cite{Tomomichel}. Entropic uncertainty relations have also proved to be of practical use in identifying security proofs of device independent QKD~\cite{Renner}. Recently Branciard et. al.~\cite{Branciard} showed for the first time that steering and security of one sided device independent (1SDI) QKD are related. In the following, we focus on the implications of joint measurability on the secred key rate in QKD against collective attacks by an adversary Eve. 

Suppose that Eve prepares a three party quantum state $\rho_{ABE}$ and gives the $A$, $B$ parts to Alice and Bob, keeping the part $E$ with her. Alice measures the observables $\mathbbm{X}$, $\mathbbm{Z}$ randomly on the state she receives  and Bob tries to predict  Alice's results by his measurements  $\mathbbm{X}'$, $\mathbbm{Z}'$. In order to generate a key,  Alice communicates her choice of measurements to Bob. Even if this communication is overheard by Eve,   Alice and Bob can generate a secure key --  provided the correlations between their measurement outcomes fare better than those between Eve and Alice. More specifically, if the difference between the mutual informations $S(\mathbbm{X}: B)=\left(\rho_{A}^\mathbbm{X}\right)+S(B)-S\left(\rho_{AB}^\mathbbm{X}\right)=S\left(\rho_{A}^\mathbbm{X}\right)
-S(\mathbbm{X}\vert B)$ and $S(\mathbbm{X}: E)=S\left(\rho_{A}^\mathbbm{X}\right)+S(E)-S\left(\rho_{AE}^\mathbbm{X}\right)=S\left(\rho_{A}^\mathbbm{X}\right)-S(\mathbbm{X}\vert E)$ (corresponding to the measurement of $\mathbbm{X}$ at Alice's end) is positive, Alice and Bob can always generate a secure key. 

The amount of key $K$ that Alice and Bob can generate per state is lower bounded by~\cite{Winter} 
\begin{equation}
\label{kr} 
K\geq S(\mathbbm{X}: B)- S(\mathbbm{X}: E)=S(\mathbbm{X}\vert E)-S(\mathbbm{X}\vert B)
\end{equation} 

It may be noted that when Alice's measurement outcomes of $\mathbbm{X}$, $\mathbbm{Z}$ are simultaneously stored in the quantum memories of Eve and Bob respectively, the following trade-off relation for the entropies $S(\mathbbm{X}\vert E)$, $S(\mathbbm{Z}\vert B)$  ensues~\cite{Berta, RB, Coles}: 
\begin{equation}
\label{evebound}
 S(\mathbbm{X}\vert E)+S(\mathbbm{Z}\vert B)\geq -2\log_2 {\cal C}(\mathbbm{X},\mathbbm{Z})
\end{equation}
And, employing (\ref{evebound}) in (\ref{kr}), one obtains 
\begin{eqnarray}
\label{kr2} 
K&\geq& S(\mathbbm{X}\vert E)+S(\mathbbm{Z}\vert B)-[S(\mathbbm{X}\vert B)+S(\mathbbm{Z}\vert B)]\nonumber \\
&\geq& -2\log_2 {\cal C}(\mathbbm{X},\mathbbm{Z})- [S(\mathbbm{X}\vert B)+S(\mathbbm{Z}\vert B)]
\end{eqnarray} 
As $H(\mathbbm{X}\vert \mathbbm{X}')\geq S(\mathbbm{X}\vert B)$, $H(\mathbbm{Z}\vert 
\mathbbm{Z}')\geq S(\mathbbm{Z}\vert B)$,    the lower bound of the inequality (\ref{kr2}) can be simplified to obtain~\cite{Berta},   
\begin{eqnarray}
\label{kr3} 
K\geq -2\log_2 {\cal C}(\mathbbm{X},\mathbbm{Z})- [H(\mathbbm{X}\vert \mathbbm{X}')+H(\mathbbm{Z}\vert 
\mathbbm{Z}')].  
\end{eqnarray} 

It is clear that when Bob is constrained to perform measurement of compatible POVMs $\mathbbm{X}', \mathbbm{Z}'$, the conditional entropies $H(\mathbbm{X}\vert \mathbbm{X}'),\,  H(\mathbbm{Z}\vert\mathbbm{Z}')$ satisfy the entropic steering inequality:  $H(\mathbbm{X}\vert \mathbbm{X}')+H(\mathbbm{Z}\vert\mathbbm{Z}')\geq -2\log_2 {\cal C}(\mathbbm{X},\mathbbm{Z})$  (see (\ref{esr})), in which case the key rate is not ensured to be positive.  Bob must be equipped to perform  incompatible measurements at his end (so that it is possible to witness violation of the steering inequality by  beating the  bound  $-2\log_2 {\cal C}(\mathbbm{X},\mathbbm{Z})$ on entropic uncertainties and attain the refined bound of $-2\log_2 {\cal C}(\mathbbm{X},\mathbbm{Z})+S(A\vert B)$  as in (\ref{BobM})) in order that a positive key rate ensues. In other words,  quantum advantage for security in  QKD against collective attacks by Eve is not envisaged,  when  Bob is constrained to perform compatible measurements only.

\section{Conclusions} 

Measurement outcomes of a pair of non-commuting observables reveal a trade-off, which is quantified by uncertainty relations. Entropic uncertainty relation~\cite{MU} constrains the sum of entropies associated with the probabilities of  outcomes of a pair of observables. An extended entropic uncertainty relation~\cite{Berta} brought out that it is possible to beat the lower bound on uncertainties when the system is entangled with a quantum memory. In this paper we have explored the entropic uncertainty relation when the entangled quantum memory is restricted to record the outcomes of jointly measurable POVMs only. With this constraint on the measurements, the entropies satisfy an entropic steering inequality~\cite{Howell}. Thus, we identify that an entangled quantum memory, which is limited to record results of compatible POVMs, cannot assist in beating the entropic uncertainty bound. As a consequence, we show that the quantum advantage in ensuring security in key distribution against collective attacks is lost, even though a suitable entangled state is employed  -- but with the joint measurability constraint.


\begin{thebibliography}{0}
\bibitem{note3} The roles played by Alice and Bob in non-local steering task (where conventionally Alice is an untrusted party and Bob needs to check violation/nonviolation of a steering inequality to verify if Alice's claim -- that they share an entangled state -- is true/false) is interchanged here so as to be consistent with the convention of Ref.~\cite{Berta}.
\bibitem{note2} Note that 
$H(\mathbbm{X}\vert \mathbbm{X}')=-\sum_{x, x'}\, p(x\vert x')\, \log_2\, p(x\vert x')$ --  where 
$p(x,x')={\rm Tr}[\rho_{AB} E_{\mathbbm{X}}(x)\otimes E_{\mathbbm{X}'}(x')],\ p(x\vert x')=p(x, x')/p(x')$ --   is the conditional Shannon entropy associated with the probabilities of Alice finding the outcome $x$ of the POVM $\mathbbm{X}$, when Bob has obtained the outcome $x'$ in the measurement of  $\mathbbm{X}'$ and 
$p(x')={\rm Tr}[\rho_{B}\,E_{\mathbbm{X}'}(x')]=\sum_{x}\, p(x,x')$ is the probability of Bob's outcome $x'$ in the measurement of $\mathbbm{X}'$.
\end{thebibliography}
\end{document}